\documentclass{article}
\usepackage{spconf,amsmath,graphicx,hyperref,booktabs,multirow,xcolor,array,arydshln}

\usepackage{adjustbox}
\usepackage{array}
\usepackage{tabularx}
\usepackage{minted}
\usepackage[most]{tcolorbox}
\usepackage{enumitem}
\usepackage{multicol}
\usepackage{listings}
\usepackage{float}

\title{AURA Score: A Metric For Holistic Audio Question Answering Evaluation}
\name{Satvik Dixit, Soham Deshmukh, Bhiksha Raj}
\address{Carnegie Mellon University}
\begin{document}
%
\maketitle

\begin{abstract}
Audio Question Answering (AQA) is a key task for evaluating Audio-Language Models (ALMs), yet assessing open-ended responses remains challenging. Existing metrics used for AQA such as BLEU, METEOR and BERTScore, mostly adapted from NLP and audio captioning, rely on surface similarity and fail to account for question context, reasoning, and partial correctness. To address the gap in literature, we make three contributions in this work. First, we introduce AQEval to enable systematic benchmarking of AQA metrics. It is the first benchmark of its kind, consisting of 10k model responses annotated by multiple humans for their correctness and relevance. Second, we conduct a comprehensive analysis of existing AQA metrics on AQEval, highlighting weak correlation with human judgment, especially for longer answers. Third, we propose a new metric - AURA score, to better evaluate open-ended model responses. On AQEval, AURA achieves state-of-the-art correlation with human ratings, significantly outperforming all baselines.  
Through this work, we aim to highlight the limitations of current AQA evaluation methods and motivate better metrics. We release both the AQEval benchmark and the AURA metric to support future research in holistic AQA evaluation.\footnote{\url{https://github.com/satvik-dixit/AURA}}
\end{abstract}
\vspace{-0.05 in}

\vspace{-0.05 in}
\section{Introduction}
\label{introduction}
\vspace{-0.05 in}
The rapid advancement of ALMs has significantly enhanced our ability to analyze and reason about audio using natural language. These models jointly process audio and text inputs to generate free-form textual responses. As ALMs continue to scale with larger training datasets \cite{ltu, ltuas, mellow, sdthesis} and parameter counts \cite{qwenaudio, ghosh2024gama, qwenaudio2}, they demonstrate remarkable capabilities across diverse audio understanding tasks. Importantly, ALMs are moving beyond simple closed-ended classification toward more complex open-ended tasks.

\begin{figure}
    \centering
    \includegraphics[width=1\linewidth]{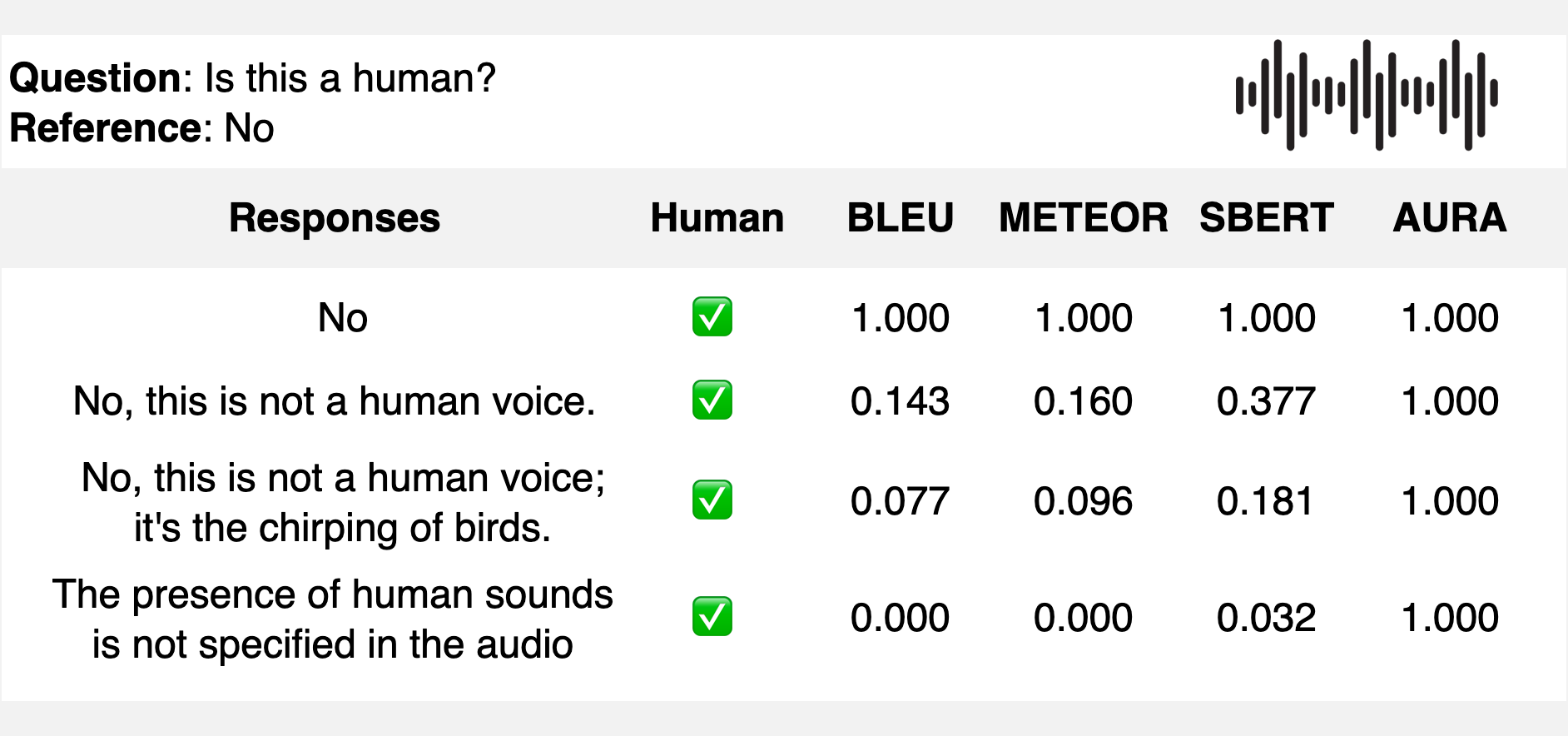}
    \vspace{-0.15 in}
    \caption{\textbf{Examples of metrics failing at AQA evaluation.} 
    As the response gets more complex, traditional metrics struggle.\vspace{-0.05 in}}
    \label{fig:intro}
\end{figure}

Open-ended audio question answering (AQA) requires models to answer natural language questions about an audio clip without being constrained to predefined options. This shift toward free-form answers introduces a major challenge for evaluation. Traditional classification metrics like accuracy and F1 score are inapplicable, leading researchers to borrow text-based NLG metrics such as BLEU \cite{bleu}, ROUGE \cite{rouge}, METEOR \cite{meteor} and BERTScore \cite{zhang2019bertscore}. While these capture lexical overlap or embedding similarity, they fail to assess whether an answer is contextually correct given the question. As shown in Figure~\ref{fig:intro}, this often results in misleading evaluations, especially for complex responses.

Automated Audio Captioning (AAC) has inspired specialized metrics such as FENSE \cite{fense} and MACE \cite{dixit2024mace}, which go beyond string matching by leveraging embeddings. However, these remain question-agnostic: they judge whether a caption matches an audio clip, not whether a response correctly addresses a specific question about the clip. Consequently, current evaluation methods misalign with human judgment, especially for nuanced, partially correct, or long responses. Humans, by contrast, naturally reason about the relationship between the question and response: when asked “Is there a dog barking?”, we treat “Yes” and “A dog is barking” as equally correct, and can also recognize partially correct answers. Effective evaluation metrics must therefore incorporate both contextual understanding and reasoning.

Our main contributions are: \vspace{-0.1in}
\begin{itemize}[noitemsep]
    \item AQEval, a new benchmark dataset for evaluating AQA metrics using human judgments. AQEval contains model-responses to audio-based questions, each annotated by five human raters for correctness, enabling systematic analysis of metric alignment with human judgments.
    \item A systematic study of existing metrics and their limitations. Our evaluation reveals that widely used metrics correlate weakly with human preference, particularly for longer or more complex answers.
    \item 
    AURA (Audio Response Assessment) Score, a metric that leverages the reasoning abilities of LLMs to evaluate correctness given the question and reference. On AQEval, AURA achieves significantly higher correlation with human ratings than prior metrics. We further conduct ablation studies to analyze the role of prompting strategies, few-shot examples, rationalization techniques, and choice of LLM.
\end{itemize}

\vspace{-0.05 in}
\section{AQEval} \label{sec:aqeval}
\vspace{-0.05 in}
We construct AQEval to evaluate the effectiveness of metrics in aligning with human judgments. The dataset is built from fixed audio files and includes diverse model responses, with details on construction, response collection, human annotation, and composition provided below.

\vspace{-0.1in}
\subsection{Data construction} \label{subsec: data collection}
\vspace{-0.05in}

\noindent\textbf{Datasets.} AQEval combines two widely used AQA datasets: ClothoAQA \cite{lipping2022clotho} and OpenAQA \cite{gong2023listen}, selected for their diversity in question types and reference formats. ClothoAQA provides binary and single-word QA from human annotators, while OpenAQA includes long-form responses generated by LLMs. For ClothoAQA, we retain only examples with majority annotator agreement, randomly sampling 500 binary and 500 single-word pairs (1,000 total). For OpenAQA, we use subsets derived from Clotho \cite{drossos2020clotho} and AudioCaps \cite{kim2019audiocaps}, filtering out ambiguous cases to retain 1,500 pairs. Together, this yields a broad benchmark spanning short to long-form AQA. Examples of question types are shown in Appendix Figure~\ref{fig:QTypes}.

\vspace{-0.1in}
\begin{table}[ht!]
\centering
\caption{AQEval composition with type representing the split between binary and word types for Clotho and between short, medium and long for OpenAQA}
\begin{adjustbox}{width=0.45\textwidth}
\begin{tabular}{lcccc}
\toprule
 Source & \# QA  & Q Vocab & A Vocab & Type \\
\midrule
 Clotho AQA test & 1327 & 413 & 116 &  665 / 662\\
 Clotho AQA val & 1537 & 474 & 100 &  781 / 756\\
 Clotho AQA train & 1767 & 477 & 108 &  918 / 849\\
\midrule
 AudioCaps train & 1454 & 591 & 1358 &  498 / 488 / 468\\
 AudioCaps val & 1313 & 563 & 1304 & 451 / 438 / 424\\
 Clotho train & 1373 & 666 & 1358 &  487 / 483 / 403\\
 Clotho val & 1203 & 557 & 1177 &  437 / 420 / 346\\
\bottomrule
\end{tabular}
\end{adjustbox}
\vspace{-0.05 in}
\label{tab:data_composition}
\end{table}

\noindent \textbf{Model Responses.} To generate candidate answers for our dataset, we use a variety of ALMs, including Qwen Audio-Chat \cite{chu2023qwen}, Audio Flamingo \cite{kong2024audio}, GAMA \cite{ghosh2024gama}, and Qwen2 Audio  \cite{chu2024qwen2}. These models were selected based on their popularity, public availability, and architectural diversity. For each question, responses are gathered from all four models, resulting in 10k (question, reference, response) triplets.

\noindent\textbf{Human Annotation.} We obtained large-scale crowdsourced annotations via Amazon Mechanical Turk (MTurk). The 10k examples were split into 8k for testing and 2k for validation. Each answer was rated by 5 annotators on a binary scale (correct/incorrect). The annotation interface and further details on worker qualifications are provided in Appendix Section~\ref{sec:human_annotation} and Figure~\ref{fig:mturk_setup}. We aggregate scores as: 1.0 if 4–5 raters marked correct, 0.5 if 2–3 marked it as correct, and 0.0 otherwise, following \cite{manas2024improving}. This scheme captures partial correctness, addressing ambiguity in open-ended AQA. After manual filtering, the final dataset contains 9,974 entries.

\noindent\textbf{Data Composition.} AQEval is built from Clotho \cite{drossos2020clotho} and AudioCaps \cite{kim2019audiocaps} audio files with paired questions, references, model responses, and annotations. The full composition of training and validation splits is shown in Table~\ref{tab:data_composition}.

\vspace{-0.05 in}
\section{AURA: Audio Response Assessment}
\vspace{-0.05 in}

\begin{figure}
    \centering
    \includegraphics[width=1\linewidth]{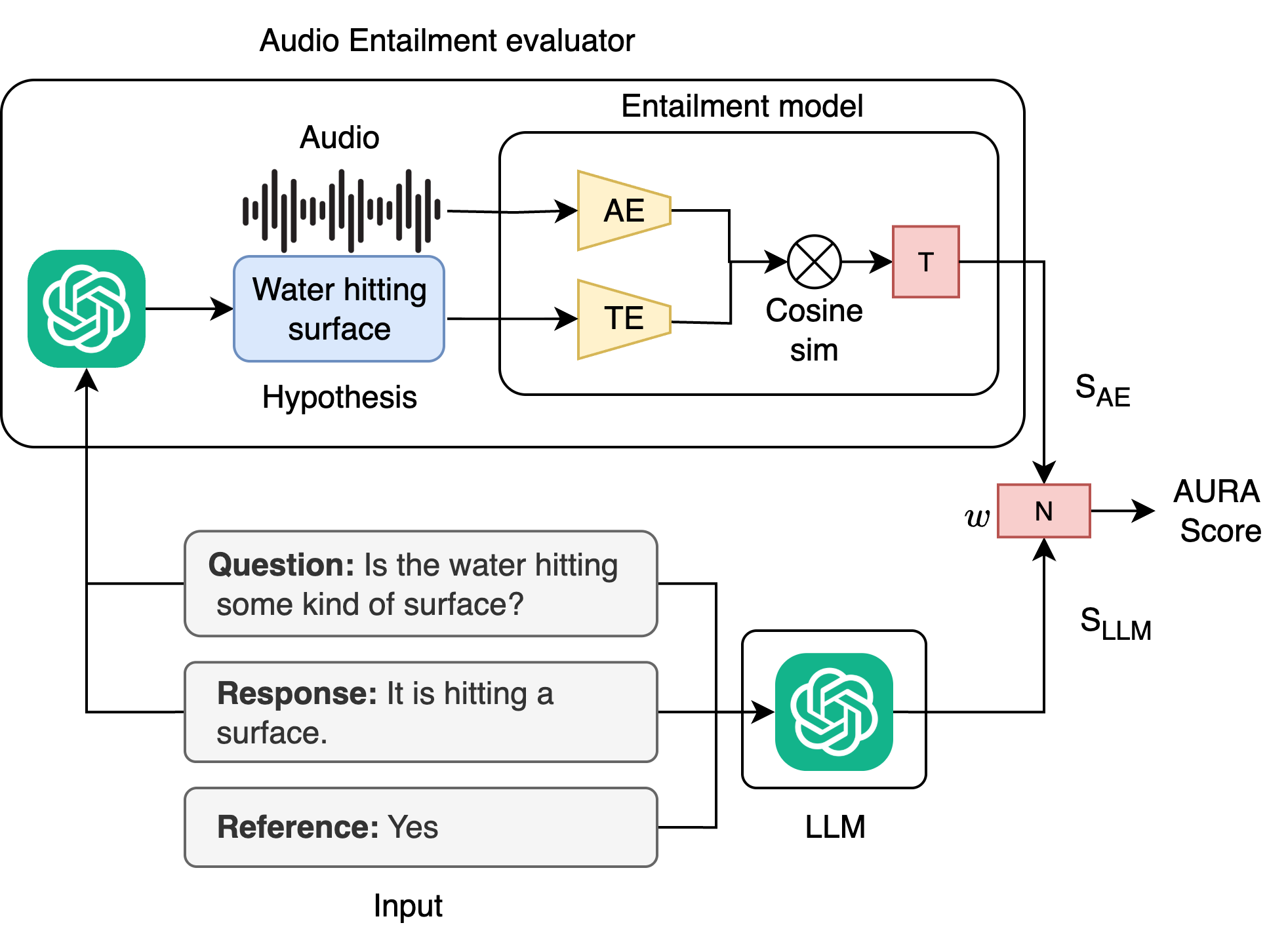}
    \vspace{-0.2in}
    \caption{\textbf{Method overview.} The AURA metric evaluates a response given the audio, question and reference. The LLM reformulates the question and response into a hypothesis and the entailment model determines if the audio entails the hypothesis. This Audio Entailment score is combined with an LLM-based correctness score as a weighted sum followed by normalization (shown at N) to get the AURA score. \vspace{-0.2in}}
    \label{fig:AURA_method}
\end{figure}

We introduce AURA, a novel evaluation metric for open-ended AQA which evaluates the quality of a model-generated response. To evaluate the correctness of a response in context, AURA leverages a large language model (LLM) with few-shot in-context learning along with an audio entailment component. AURA jointly evaluates textual correctness and audio grounding of model responses as shown in Figure \ref{fig:AURA_method}. 

Let $q$ be the question about audio $a$, $ref$ be the reference answer and $r$ be the model-generated response. Then AURA score is computed by combining two components:

\noindent\textbf{LLM-based scoring.} To assess correctness in context, AURA leverages a large language model (LLM) with few-shot in-context learning. The LLM is prompted with $q$, $r$ and $ref$, and outputs a categorical rating. Each prompt includes a task description, several annotated demonstrations (few-shot examples), and a target test instance. Demonstrations consist of a question, reference, candidate response, rating, and a natural language rationale. The LLM is instructed to first generate a rationale, then provide a rating on a three-point scale (1 = incorrect, 2 = ambiguous/partially correct, 3 = correct), following \cite{manas2024improving}. This process encourages transparent reasoning. The resulting score $S_{\text{LLM}}$ is mapped to $0, 0.5$ and $1$ and captures contextual correctness. The full prompt structure, including few-shot examples and CoT instructions, is available in Appendix Section~\ref{sec:prompts} (Figures~\ref{fig:evaluation_prompt} and \ref{fig:evaluation_examples}).

\noindent\textbf{Audio entailment.} To ensure responses are grounded in the audio, AURA incorporates an \textit{audio entailment} check \cite{audioentail}. The question $q$ and response $r$ are reformulated into a declarative hypothesis $h$ (e.g., \textit{“Is there a dog barking?” + “Yes”} $\rightarrow$ \textit{“A dog is barking.”}). This reformulation is guided by the LLM prompt shown in Appendix Figure~\ref{fig:hypothesis_prompt}. The task reduces to determining whether the audio $a$ entails the hypothesis $h$. Here, as the audio entailment model, we use CLAP. Specifically, we compute cosine similarity ($\cos(\cdot, \cdot)$) between the CLAP audio embedding ($E_a(a)$) of the audio and CLAP text embeddings ($E_t(h)$)embeddings of the hypothesis:
\begin{align}
s &= \cos(E_a(a), E_t(h)),
\end{align}
Thresholded similarity (shown as T in Figure \ref{fig:AURA_method}) classifies the response as entailment (+1), neutral (0), or contradiction (–1), yielding the audio entailment score $S_{\text{AE}}$.  

\noindent\textbf{AURA score.} The components are combined as follows:
\begin{equation}
\text{AURA}(q, a, r, ref) = Normalised(S_{\text{LLM}} + w \cdot S_{\text{AE}}),
\end{equation}
where $w \geq 0$ balances the contribution of audio entailment and Normalised refers to min-max scaling to get the final AURA score as a value between 0 and 1.

\vspace{-0.05 in}
\section{Experimental setup} \label{sec: experimental setup}
\vspace{-0.05 in}

\noindent \textbf{Dataset.} 
All experiments are conducted on the proposed AQEval benchmark, which comprises 8k examples, each containing an audio clip, a question, a reference answer, a model-generated response, and corresponding human annotations. The dataset details can be found in Section \ref{subsec: data collection}.

\noindent \textbf{Task Setup.}
To assess metric quality, we compute each metric’s score for every example in AQEval and evaluate its alignment with human judgments using Pearson’s rank correlation coefficient ($\rho$). For each experimental setup, we boldface the highest-performing metric.

\noindent \textbf{Baselines.} 
We compare AURA against a wide range of baselines. Traditional NLP metrics include BLEU \cite{papineni2002bleu}, ROUGE-L \cite{lin2004rouge}, METEOR \cite{agarwal2008meteor}, and CIDEr \cite{vedantam2015cider}, which measure $n$-gram overlap. SPICE \cite{anderson2016spice} and SPIDEr \cite{liu2017improved} extend this approach by incorporating object graphs and were originally developed for image captioning. More recent metrics such as FENSE \cite{zhou2022can} and MACE \cite{dixit2024mace} leverage embedding similarity (via SBERT and CLAP, respectively) and were designed specifically for audio captioning. 

\vspace{-0.05 in}
\section{Results}
\vspace{-0.05 in}

We evaluate different AQA metrics on the AQEval benchmark. We also benchmark metrics across question types.

\noindent\textbf{Benchmarking Evaluation Metrics.} 
Table \ref{tab:evaluation_metrics} presents the correlations between human judgments and various AQA metrics, including AURA and baseline metrics. We further segregate the results based on the underlying model response and the dataset AQA pair is sourced from. Our results demonstrate that AURA consistently achieves the highest correlation with human judgments, significantly outperforming traditional metrics across most datasets and models. It even outperforms the LLM baseline by 16.02\% on ClothoAQA overall and by 4.31\% on OpenAQA overall.

\begin{table}[!ht]
\centering
\vspace{-0.1in}
\caption{Correlation between metrics and human judgments on AQEval across question types.}
\begin{adjustbox}{width=0.45\textwidth}
\begin{tabular}{lclllcc}
\toprule
Metric          & Binary  &Word  &Short  &Medium & Long & Overall \\
\midrule
BLEU  & 10.92  &36.92  &32.86  &10.98 & 17.02 & 23.91 \\
METEOR & 12.34  &39.58  &36.56  &15.91 & 18.22 & 27.86 \\
ROUGE L & 12.43  &43.25  &38.67  &8.92 & 15.19 & 27.34 \\
CIDER & 9.96  &38.89  &29.77  &6.51 & 10.99 & 19.42 \\
SPICE & -0.93  &43.46  &31.33  &4.99 & 12.10 & 20.24 \\
SPIDER & 8.04  &42.51  &31.20  &6.56 & 12.16 & 21.22 \\
FENSE & 37.11  &47.07  &29.45  &27.78 & 19.43 & 17.52 \\
MACE & 16.99  &39.75  &31.53  &20.63 & 19.42 & 20.30 \\
\midrule
LLM & 67.02  &57.47  &47.23  &50.69 & 39.42 & 56.64 \\
AURA & 81.20  &64.65  &46.60  &53.12 & 42.03 & 61.80 \\
\bottomrule
\end{tabular}
\end{adjustbox}
\vspace{-0.1in}
\label{tab:single_column_metrics}
\end{table}

\begin{table*}[ht!]
\centering
\caption{Correlation between evaluation metrics and human judgments on AQEval}
\begin{adjustbox}{width=\textwidth}
\begin{tabular}{lcccccccccc}
\toprule
Metric          & \multicolumn{2}{c}{Audio Flamingo} & \multicolumn{2}{c}{GAMA} & \multicolumn{2}{c}{Qwen Audio 1} & \multicolumn{2}{c}{Qwen Audio 2} & \multicolumn{2}{c}{Overall} \\
\cmidrule(lr){2-3} \cmidrule(lr){4-5} \cmidrule(lr){6-7} \cmidrule(lr){8-9} \cmidrule(lr){10-11}
                & ClothoAQA & OpenAQA & ClothoAQA & OpenAQA & ClothoAQA & OpenAQA & ClothoAQA & OpenAQA & ClothoAQA & OpenAQA \\
\midrule
BLEU  & 40.88 & 16.72 & 31.58 & 17.26 & 26.59 & 15.64 & 24.93 & 18.57 & 31.00 & 17.05 \\
METEOR  & 40.31 & 23.76 & 28.93 & 21.92 & 31.17 & 22.28 & 26.19 & 22.58 & 31.65 & 22.64 \\
ROUGE L  & 40.88 & 25.86 & 31.99 & 15.40 & 32.15 & 15.33 & 29.93 & 19.66 & 33.74 & 19.06 \\
CIDER  & 40.88 & 17.87 & 15.50 & 14.47 & 25.44 & 10.67 & 26.54 & 15.98 & 27.09 & 14.75 \\
SPICE  & 25.43 & 15.71 & 14.43 & 11.73 & 18.76 & 12.77 & 21.83 & 15.24 & 20.11 & 13.86 \\
SPIDER  & 40.26 & 18.79 & 19.19 & 15.07 & 26.53 & 11.29 & 27.35 & 16.69 & 28.33 & 15.46 \\
MACE & 35.37 & 17.82 & 12.14 & 22.30 & 18.34 & 19.12 & 19.47 & 19.61 & 21.33 & 19.71 \\
FENSE & 24.57 & 12.00 & 14.07 & 31.79 & 32.97 & 18.36 & 23.31 & 24.91 & 23.73 & 21.76 \\
\midrule
LLM & 82.09 & 37.15 & 52.98 & 50.21 & 60.93 & 43.08 & 54.35 & 43.80 & 62.59 & 43.56 \\
AURA & 84.91 & 36.11 & 65.99 & 53.39 & 70.99 & 46.00 & 68.59 & 46.26 & 72.62 & 45.44 \\
\bottomrule
\end{tabular}
\end{adjustbox}
\vspace{-0.1in}
\label{tab:evaluation_metrics}
\end{table*}

\noindent\textbf{Comparing Metrics Across Question Types.} To assess the effectiveness of evaluation metrics across different types of responses, we categorize model responses on the OpenAQA questions into distinct types based on length: short, medium, and long responses. This was done by sorting the responses by length and splitting them into three equal buckets. We also show the results for the two different types of questions in the ClothoAQA dataset - binary and single-word responses. This categorization ensures a diverse set of examples to evaluate the performance of metrics across varying response lengths and complexities. For each question category, we calculate the correlation between metric scores and human judgments. The results, presented in Table \ref{tab:single_column_metrics}, reveal a trend: As the responses become longer and more complex, traditional metrics, such as BLEU, ROUGE, and METEOR, show a notable decrease in their correlation with human scores. For binary questions, the reference answers are typically 'yes' or 'no'. However, model responses are often more descriptive. This mismatch leads to poor scores for traditional n-gram-based metrics, whereas LLM-based approaches like AURA achieve high correlation. In general, traditional metrics achieve higher scores on Word and Short answer categories compared to Medium and Long ones. This highlights their inadequacy in capturing the semantic correctness of longer, more complex answers. Appendix Table~\ref{tab:qualitative_results} provides several qualitative examples that illustrate these specific failure cases. In contrast, AURA shows strong performance in all categories, even surpassing the LLM baseline by 9.1\% overall, highlighting its robustness in handling diverse types of responses.

\vspace{-0.05 in}
\section{Ablations}
\vspace{-0.05 in}
We conduct a series of ablation studies to evaluate the impact of various design choices on AURA's performance. In particular, we examine the effects of number of examples in few-shot setting, chain of thought prompting and LLM choice. The results are shown in Table \ref{tab:ablations}. The key observations are:

\noindent\textbf{Number of Demonstrations.} We assess the effect of providing multiple examples to the LLM. Demonstrations include a question, a set of reference answers, a candidate answer, an explanation, and a corresponding rating. Performance improves steadily from zero-shot to three-shot prompting, which is consistent with other studies \cite{manas2024improving, dixit2024vision}. We also find that relative gains from additional examples plateau.

\noindent\textbf{Rationalization.} We measure the impact of requiring the LLM to generate an explanation before assigning a rating to a candidate answer. We find that including this step not only improves the interpretability of the ratings but also leads to an increase in correlation with human judgments. This aligns with findings in related tasks \cite{manas2024improving}, where rationalization enhances performance by grounding the decision-making.

\noindent\textbf{Choice of LLM.} We evaluate the effect of using other LLMs, specifically Gemini (Gemini 2.5 Pro), Claude (Claude sonnet 3.5), GPT-4 (GPT-4o) instead of Llama (Llama 3.1-8B). Across the board, stronger models yield higher correlation with human judgments with particularly strong gains on long answers compared to llama. Specifically, we see about 6.01\% improvement in correlation using GPT-4o (65.88) over Llama (62.14). These results highlight that the underlying reasoning ability of the LLM is critical and that as models get better, the AURA metric will yield better results.

\begin{figure}
    \centering
    \includegraphics[width=0.9\linewidth]{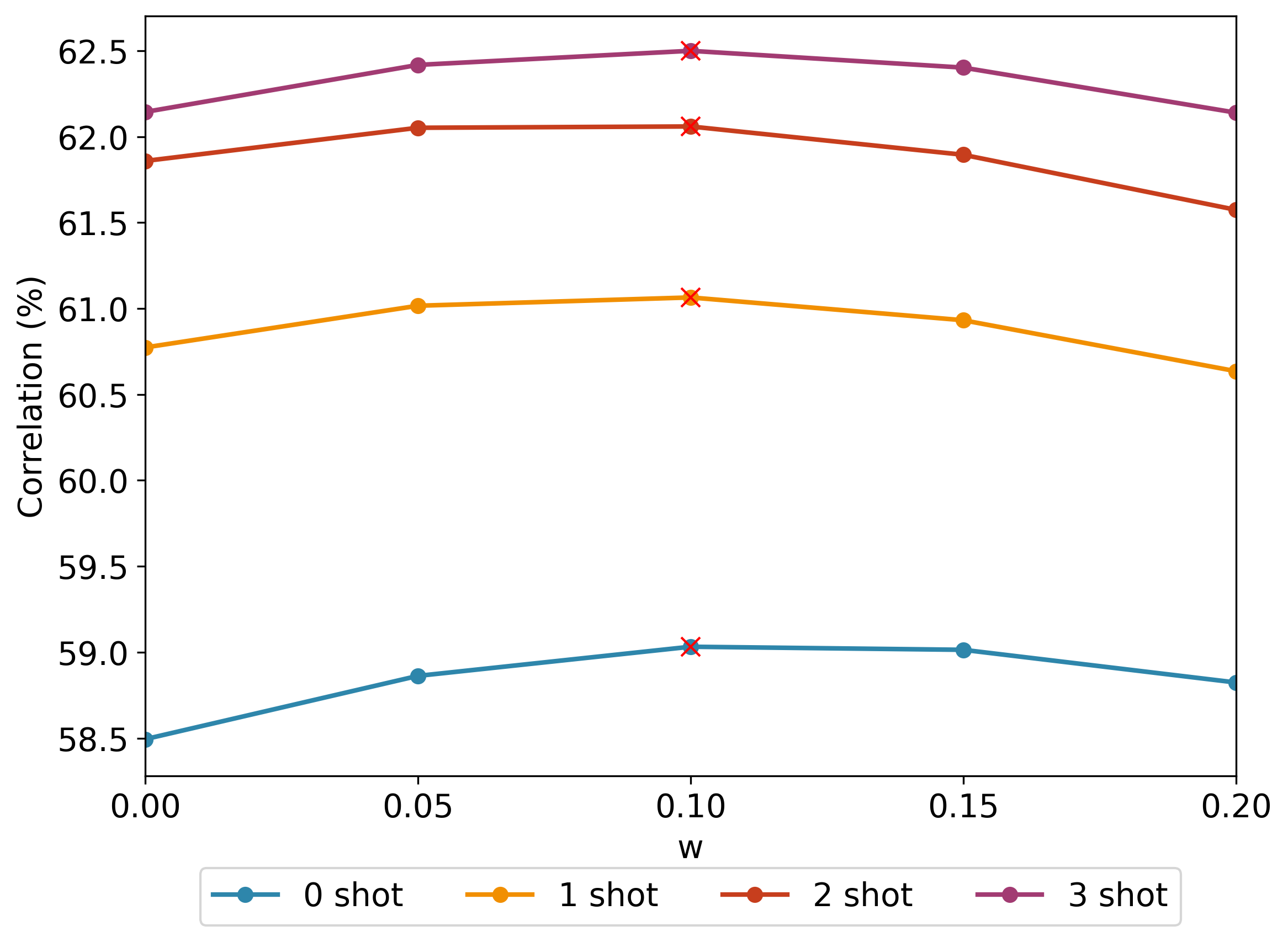}
    \caption{Effect of adding the audio entailment term.}  
    \label{fig:AURA_cot_normal_subplots}
    \vspace{-0.2in}
\end{figure}

\noindent\textbf{Audio Entailment Component.} We evaluate the effect of adding the entailment term to 1-shot, 2-shot and 3-shot LLM scores from the baseline Gemini model. We find that the correlation increases consistently for a range of weight values as shown in Figure \ref{fig:AURA_cot_normal_subplots} however the gains are small. Our current audio entailment system is a zero-shot system with about 50\% accuracy on Audio Entailment benchmark \cite{audioentail}. We believe that improvements in audio entailment detection will translate to a stronger correlation between the AURA score and human judgments.

\vspace{-0.05in}
\begin{table}[ht!]
\centering
\caption{Correlation between metrics and human judgments for different configurations of AURA on AQEval.}
\begin{adjustbox}{width=0.47\textwidth}
\begin{tabular}{lcccccc}
\toprule
Model Variant & Binary & Word & Short & Medium & Long & Overall \\
\midrule
Baseline & 57.73 & 64.59 & 43.84 & 47.34 & 46.59 & 57.97 \\
\midrule
CoT & 60.11 & 64.27 & 46.00 & 47.83 & 44.39 & 58.49 \\
\midrule
1 shot & 77.17 & 68.57 & 37.17 & 45.96 & 43.16 & 60.77 \\
2 shot & 76.74 & 70.88 & 41.10 & 46.73 & 39.91 & 61.86 \\
3 shot & 76.30 & 72.37 & 39.09 & 47.83 & 40.55 & 62.14 \\
\midrule
Gemini & 80.01 & 74.13 & 38.98 & 52.92 & 46.65 & 64.20 \\
Claude & 80.53 & 72.79 & 42.80 & 51.07 & 47.00 & 64.38 \\
GPT 4o & 80.88 & 71.58 & 44.97 & 50.61 & 49.80 & 65.88 \\
\bottomrule
\end{tabular}
\end{adjustbox}
\vspace{-0.05 in}
\label{tab:ablations}
\end{table}

\vspace{-0.1 in}
\section{Conclusion}
\vspace{-0.1 in}
We introduced AQEval, the first benchmark for evaluating AQA metrics with human judgments, and showed that existing methods fall short on nuanced responses. Our proposed metric, AURA score, leverages language-based reasoning and audio entailment to achieve SotA correlation with humans. Specifically, AURA's correlation with human judgment is 2.2 times higher than that of the best existing metric (METEOR) and 9.1\% better than the baseline LLM.  We hope this work aids the development of more sophisticated metrics for audio-language tasks.

\newpage
\section{Acknowledgments}
\label{sec:acknowledgments}
This work used Bridges2 at PSC by allocation CIS220001P from the Advanced Cyberinfrastructure Coordination Ecosystem: Services \& Support (ACCESS) program, which is supported by U.S. National Science Foundation grants \#2138259, \#2138286, \#2138307, \#2137603, and \#2138296.


\small
\bibliographystyle{IEEEbib}
\bibliography{strings,refs}

\begin{thebibliography}{10}

\bibitem{ltu}
Yuan Gong, Hongyin Luo, Alexander~H. Liu, Leonid Karlinsky, and James~R. Glass,
\newblock ``Listen, think, and understand,''
\newblock in {\em The Twelfth International Conference on Learning Representations}, 2024.

\bibitem{ltuas}
Yuan Gong, Alexander~H Liu, Hongyin Luo, Leonid Karlinsky, and James Glass,
\newblock ``Joint audio and speech understanding,''
\newblock in {\em 2023 IEEE Automatic Speech Recognition and Understanding Workshop (ASRU)}. IEEE, 2023, pp. 1--8.

\bibitem{mellow}
Soham Deshmukh, Satvik Dixit, Rita Singh, and Bhiksha Raj,
\newblock ``Mellow: a small audio language model for reasoning,''
\newblock {\em arXiv preprint arXiv:2503.08540}, 2025.

\bibitem{sdthesis}
Soham Deshmukh,
\newblock {\em Learning Audio Foundation Models for Reasoning},
\newblock Ph.D. thesis, Carnegie Mellon University, 2025.

\bibitem{qwenaudio}
Yunfei Chu, Jin Xu, Xiaohuan Zhou, Qian Yang, Shiliang Zhang, Zhijie Yan, Chang Zhou, and Jingren Zhou,
\newblock ``Qwen-audio: Advancing universal audio understanding via unified large-scale audio-language models,''
\newblock {\em arXiv preprint arXiv:2311.07919}, 2023.

\bibitem{ghosh2024gama}
Sreyan Ghosh, Sonal Kumar, Ashish Seth, Chandra Kiran~Reddy Evuru, Utkarsh Tyagi, S~Sakshi, Oriol Nieto, Ramani Duraiswami, and Dinesh Manocha,
\newblock ``{GAMA}: A large audio-language model with advanced audio understanding and complex reasoning abilities,''
\newblock in {\em Proceedings of the 2024 Conference on Empirical Methods in Natural Language Processing}, Yaser Al-Onaizan, Mohit Bansal, and Yun-Nung Chen, Eds., Miami, Florida, USA, Nov. 2024, pp. 6288--6313, Association for Computational Linguistics.

\bibitem{qwenaudio2}
Yunfei Chu, Jin Xu, Qian Yang, Haojie Wei, Xipin Wei, Zhifang Guo, Yichong Leng, Yuanjun Lv, Jinzheng He, Junyang Lin, Chang Zhou, and Jingren Zhou,
\newblock ``Qwen2-audio technical report,'' 2024.

\bibitem{bleu}
Kishore Papineni, Salim Roukos, Todd Ward, and Wei-Jing Zhu,
\newblock ``Bleu: a method for automatic evaluation of machine translation,''
\newblock in {\em Proceedings of the 40th annual meeting of the Association for Computational Linguistics}, 2002, pp. 311--318.

\bibitem{rouge}
Chin-Yew Lin,
\newblock ``{ROUGE}: A package for automatic evaluation of summaries,''
\newblock in {\em Text Summarization Branches Out}, Barcelona, Spain, July 2004, pp. 74--81, Association for Computational Linguistics.

\bibitem{meteor}
Satanjeev Banerjee and Alon Lavie,
\newblock ``{METEOR}: An automatic metric for {MT} evaluation with improved correlation with human judgments,''
\newblock in {\em Proceedings of the {ACL} Workshop on Intrinsic and Extrinsic Evaluation Measures for Machine Translation and/or Summarization}, Jade Goldstein, Alon Lavie, Chin-Yew Lin, and Clare Voss, Eds., Ann Arbor, Michigan, June 2005, pp. 65--72, Association for Computational Linguistics.

\bibitem{zhang2019bertscore}
Tianyi Zhang*, Varsha Kishore*, Felix Wu*, Kilian~Q. Weinberger, and Yoav Artzi,
\newblock ``Bertscore: Evaluating text generation with bert,''
\newblock in {\em International Conference on Learning Representations}, 2020.

\bibitem{fense}
Zelin Zhou, Zhiling Zhang, Xuenan Xu, Zeyu Xie, Mengyue Wu, and Kenny~Q Zhu,
\newblock ``Can audio captions be evaluated with image caption metrics?,''
\newblock in {\em ICASSP 2022-2022 IEEE International Conference on Acoustics, Speech and Signal Processing (ICASSP)}. IEEE, 2022, pp. 981--985.

\bibitem{dixit2024mace}
Satvik Dixit, Soham Deshmukh, and Bhiksha Raj,
\newblock ``Mace: Leveraging audio for evaluating audio captioning systems,''
\newblock {\em 2025 IEEE International Conference on Acoustics, Speech, and Signal Processing Workshops (ICASSPW)}, pp. 1--5, 2024.

\bibitem{lipping2022clotho}
Samuel Lipping, Parthasaarathy Sudarsanam, Konstantinos Drossos, and Tuomas Virtanen,
\newblock ``Clotho-aqa: A crowdsourced dataset for audio question answering,''
\newblock in {\em 2022 30th European Signal Processing Conference (EUSIPCO)}. IEEE, 2022, pp. 1140--1144.

\bibitem{gong2023listen}
Yuan Gong, Hongyin Luo, Alexander~H. Liu, Leonid Karlinsky, and James~R. Glass,
\newblock ``Listen, think, and understand,''
\newblock in {\em The Twelfth International Conference on Learning Representations}, 2024.

\bibitem{drossos2020clotho}
Konstantinos Drossos, Samuel Lipping, and Tuomas Virtanen,
\newblock ``Clotho: An audio captioning dataset,''
\newblock in {\em ICASSP 2020-2020 IEEE International Conference on Acoustics, Speech and Signal Processing (ICASSP)}. IEEE, 2020, pp. 736--740.

\bibitem{kim2019audiocaps}
Chris~Dongjoo Kim, Byeongchang Kim, Hyunmin Lee, and Gunhee Kim,
\newblock ``Audiocaps: Generating captions for audios in the wild,''
\newblock in {\em Proceedings of the 2019 Conference of the North American Chapter of the Association for Computational Linguistics: Human Language Technologies, Volume 1 (Long and Short Papers)}, 2019, pp. 119--132.

\bibitem{chu2023qwen}
Yunfei Chu, Jin Xu, Xiaohuan Zhou, Qian Yang, Shiliang Zhang, Zhijie Yan, Chang Zhou, and Jingren Zhou,
\newblock ``Qwen-audio: Advancing universal audio understanding via unified large-scale audio-language models,''
\newblock {\em arXiv preprint arXiv:2311.07919}, 2023.

\bibitem{kong2024audio}
Zhifeng Kong, Arushi Goel, Rohan Badlani, Wei Ping, Rafael Valle, and Bryan Catanzaro,
\newblock ``Audio flamingo: A novel audio language model with few-shot learning and dialogue abilities,''
\newblock in {\em Proceedings of the 41st International Conference on Machine Learning}, Ruslan Salakhutdinov, Zico Kolter, Katherine Heller, Adrian Weller, Nuria Oliver, Jonathan Scarlett, and Felix Berkenkamp, Eds. 21--27 Jul 2024, vol. 235 of {\em Proceedings of Machine Learning Research}, pp. 25125--25148, PMLR.

\bibitem{chu2024qwen2}
Yunfei Chu, Jin Xu, Qian Yang, Haojie Wei, Xipin Wei, Zhifang Guo, Yichong Leng, Yuanjun Lv, Jinzheng He, Junyang Lin, et~al.,
\newblock ``Qwen2-audio technical report,''
\newblock {\em arXiv preprint arXiv:2407.10759}, 2024.

\bibitem{manas2024improving}
Oscar Ma{\~n}as, Benno Krojer, and Aishwarya Agrawal,
\newblock ``Improving automatic vqa evaluation using large language models,''
\newblock in {\em Proceedings of the AAAI Conference on Artificial Intelligence}, 2024, vol.~38, pp. 4171--4179.

\bibitem{audioentail}
Soham Deshmukh, Shuo Han, Hazim Bukhari, Benjamin Elizalde, Hannes Gamper, Rita Singh, and Bhiksha Raj,
\newblock ``Audio entailment: Assessing deductive reasoning for audio understanding,''
\newblock {\em Proceedings of the AAAI Conference on Artificial Intelligence}, vol. 39, no. 22, pp. 23769--23777, Apr. 2025.

\bibitem{papineni2002bleu}
Kishore Papineni, Salim Roukos, Todd Ward, and Wei-Jing Zhu,
\newblock ``Bleu: a method for automatic evaluation of machine translation,''
\newblock in {\em Proceedings of the 40th annual meeting of the Association for Computational Linguistics}, 2002, pp. 311--318.

\bibitem{lin2004rouge}
Chin-Yew Lin,
\newblock ``Rouge: A package for automatic evaluation of summaries,''
\newblock in {\em Text summarization branches out}, 2004, pp. 74--81.

\bibitem{agarwal2008meteor}
Abhaya Agarwal and Alon Lavie,
\newblock ``Meteor, m-bleu and m-ter: Evaluation metrics for high-correlation with human rankings of machine translation output,''
\newblock in {\em Proceedings of the Third Workshop on Statistical Machine Translation}, 2008, pp. 115--118.

\bibitem{vedantam2015cider}
Ramakrishna Vedantam, C~Lawrence~Zitnick, and Devi Parikh,
\newblock ``Cider: Consensus-based image description evaluation,''
\newblock in {\em Proceedings of the IEEE conference on computer vision and pattern recognition}, 2015, pp. 4566--4575.

\bibitem{anderson2016spice}
Peter Anderson, Basura Fernando, Mark Johnson, and Stephen Gould,
\newblock ``Spice: Semantic propositional image caption evaluation,''
\newblock in {\em Computer Vision--ECCV 2016: 14th European Conference, Amsterdam, The Netherlands, October 11-14, 2016, Proceedings, Part V 14}. Springer, 2016, pp. 382--398.

\bibitem{liu2017improved}
Siqi Liu, Zhenhai Zhu, Ning Ye, Sergio Guadarrama, and Kevin Murphy,
\newblock ``Improved image captioning via policy gradient optimization of spider,''
\newblock in {\em Proceedings of the IEEE international conference on computer vision}, 2017, pp. 873--881.

\bibitem{zhou2022can}
Zelin Zhou, Zhiling Zhang, Xuenan Xu, Zeyu Xie, Mengyue Wu, and Kenny~Q Zhu,
\newblock ``Can audio captions be evaluated with image caption metrics?,''
\newblock in {\em ICASSP 2022-2022 IEEE International Conference on Acoustics, Speech and Signal Processing (ICASSP)}. IEEE, 2022, pp. 981--985.

\bibitem{dixit2024vision}
Satvik Dixit, Laurie~M Heller, and Chris Donahue,
\newblock ``Vision language models are few-shot audio spectrogram classifiers,''
\newblock {\em arXiv preprint arXiv:2411.12058}, 2024.

\bibitem{sakshi2024mmau}
S~Sakshi, Utkarsh Tyagi, Sonal Kumar, Ashish Seth, Ramaneswaran Selvakumar, Oriol Nieto, Ramani Duraiswami, Sreyan Ghosh, and Dinesh Manocha,
\newblock ``Mmau: A massive multi-task audio understanding and reasoning benchmark,''
\newblock {\em arXiv preprint arXiv:2410.19168}, 2024.

\bibitem{ma2025mmar}
Ziyang Ma, Yinghao Ma, Yanqiao Zhu, Chen Yang, Yi-Wen Chao, Ruiyang Xu, Wenxi Chen, Yuanzhe Chen, Zhuo Chen, Jian Cong, et~al.,
\newblock ``Mmar: A challenging benchmark for deep reasoning in speech, audio, music, and their mix,''
\newblock {\em arXiv preprint arXiv:2505.13032}, 2025.

\bibitem{behera2023aquallm}
Swarup~Ranjan Behera, Krishna~Mohan Injeti, Jaya Sai~Kiran Patibandla, Praveen~Kumar Pokala, and Balakrishna~Reddy Pailla,
\newblock ``Aquallm: audio question answering data generation using large language models,''
\newblock {\em arXiv preprint arXiv:2312.17343}, 2023.

\bibitem{yang2025multi}
Chao-Han~Huck Yang, Sreyan Ghosh, Qing Wang, Jaeyeon Kim, Hengyi Hong, Sonal Kumar, Guirui Zhong, Zhifeng Kong, S~Sakshi, Vaibhavi Lokegaonkar, et~al.,
\newblock ``Multi-domain audio question answering toward acoustic content reasoning in the dcase 2025 challenge,''
\newblock {\em arXiv preprint arXiv:2505.07365}, 2025.

\bibitem{sudarsanam2023attention}
Parthasaarathy Sudarsanam and Tuomas Virtanen,
\newblock ``Attention-based methods for audio question answering,''
\newblock in {\em 2023 31st European Signal Processing Conference (EUSIPCO)}. IEEE, 2023, pp. 750--754.

\bibitem{behera2023towards}
Swarup~Ranjan Behera, Pailla~Balakrishna Reddy, Achyut~Mani Tripathi, Megavath~Bharadwaj Rathod, and Tejesh Karavadi,
\newblock ``Towards multi-lingual audio question answering,''
\newblock in {\em Proceedings of INTERSPEECH 2023}, 2023, pp. 356--360.

\bibitem{wang2025careaqa}
Tsai-Ning Wang, Lin-Lin Chen, Neil Zeghidour, and Aaqib Saeed,
\newblock ``Careaqa: A cardiac and respiratory audio question answering model for open-ended diagnostic reasoning,''
\newblock {\em arXiv preprint arXiv:2505.01199}, 2025.

\bibitem{liu2024music}
Shansong Liu, Atin~Sakkeer Hussain, Chenshuo Sun, and Ying Shan,
\newblock ``Music understanding llama: Advancing text-to-music generation with question answering and captioning,''
\newblock in {\em ICASSP 2024-2024 IEEE International Conference on Acoustics, Speech and Signal Processing (ICASSP)}. IEEE, 2024, pp. 286--290.

\bibitem{gontier2023spice+}
F{\'e}lix Gontier, Romain Serizel, and Christophe Cerisara,
\newblock ``Spice+: Evaluation of automatic audio captioning systems with pre-trained language models,''
\newblock in {\em ICASSP 2023-2023 IEEE International Conference on Acoustics, Speech and Signal Processing (ICASSP)}. IEEE, 2023, pp. 1--5.

\bibitem{wijngaard2023aces}
Gijs Wijngaard, Elia Formisano, Bruno~L Giordano, and Michel Dumontier,
\newblock ``Aces: Evaluating automated audio captioning models on the semantics of sounds,''
\newblock in {\em 2023 31st European Signal Processing Conference (EUSIPCO)}. IEEE, 2023, pp. 770--774.

\bibitem{kocmi2023large}
Tom Kocmi and Christian Federmann,
\newblock ``Large language models are state-of-the-art evaluators of translation quality,''
\newblock {\em arXiv preprint arXiv:2302.14520}, 2023.

\bibitem{liu2023g}
Yang Liu, Dan Iter, Yichong Xu, Shuohang Wang, Ruochen Xu, and Chenguang Zhu,
\newblock ``G-eval: Nlg evaluation using gpt-4 with better human alignment,''
\newblock {\em arXiv preprint arXiv:2303.16634}, 2023.

\bibitem{zheng2023judging}
Lianmin Zheng, Wei-Lin Chiang, Ying Sheng, Siyuan Zhuang, Zhanghao Wu, Yonghao Zhuang, Zi~Lin, Zhuohan Li, Dacheng Li, Eric Xing, et~al.,
\newblock ``Judging llm-as-a-judge with mt-bench and chatbot arena,''
\newblock {\em Advances in Neural Information Processing Systems}, vol. 36, pp. 46595--46623, 2023.

\bibitem{lin2023llm}
Yen-Ting Lin and Yun-Nung Chen,
\newblock ``Llm-eval: Unified multi-dimensional automatic evaluation for open-domain conversations with large language models,''
\newblock {\em arXiv preprint arXiv:2305.13711}, 2023.

\end{thebibliography}

\newpage
\normalsize
\appendix

\begin{figure*}
    \centering
    \includegraphics[width=1\linewidth]{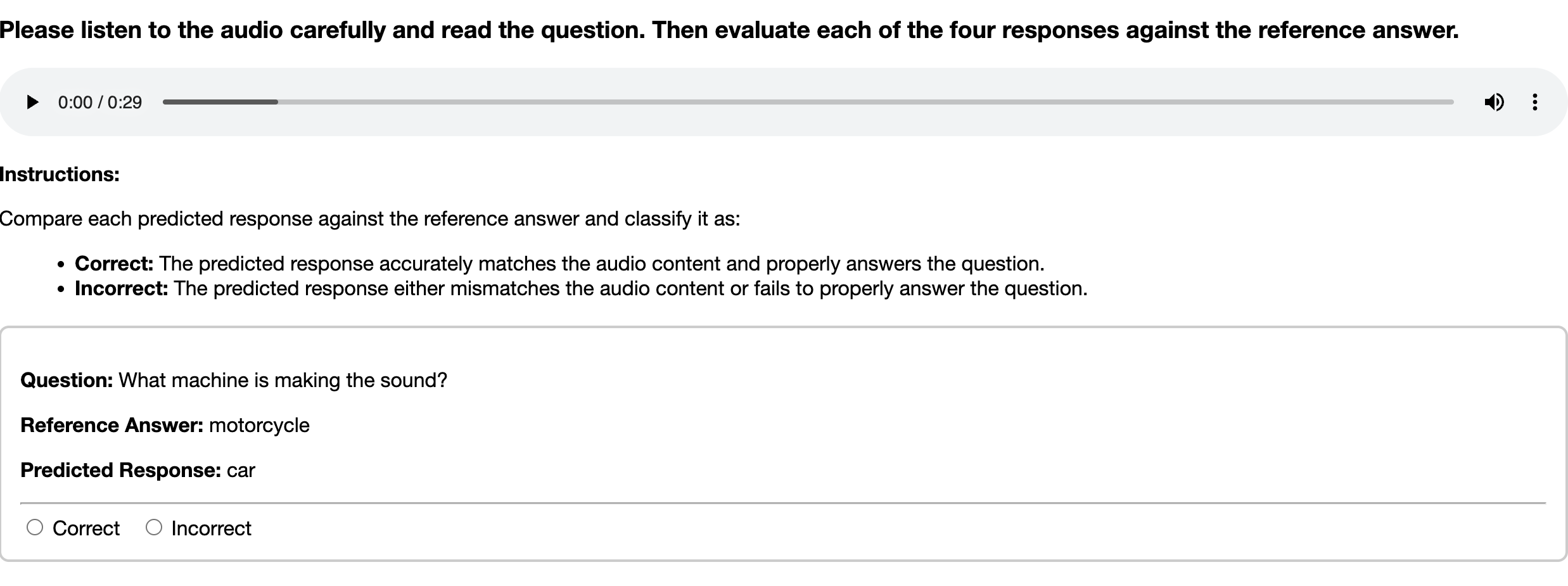}
    \caption{\textbf{MTurk interface used for our human annotation.} Annotators listened to an audio clip, reviewed the question and reference answer, and provided a binary correctness judgment (Correct/Incorrect) on the candidate response}
    \label{fig:mturk_setup}
\end{figure*}

\section{Related work}

\noindent\textbf{NLP-based Metrics.}
In many prior works, model responses are evaluated by converting open-ended tasks into close-ended multiple-choice formats where one can simply match against a known answer option \cite{sakshi2024mmau, ma2025mmar} or by using string matching accuracy for open-ended reponses \cite{behera2023aquallm, yang2025multi, sudarsanam2023attention}. To move beyond classification-style evaluation, researchers have employed natural language generation (NLG) metrics such as BLEU \cite{bleu}, ROUGE \cite{rouge} and METEOR \cite{meteor}. BLEU and ROUGE rely on exact n-gram overlaps between generated responses and references, which often leads to harsh penalties for valid paraphrases. METEOR improves upon this by incorporating stemming and synonym matching. These metrics have been adopted for AQA evaluation in recent work \cite{behera2023towards, wang2025careaqa, liu2024music}. However, none of these metrics were designed with audio content in mind, and they fundamentally lack the capacity to assess whether a response is grounded in the acoustic signal. For example, the sounds of “a horse trotting” and “knocking on a table” may be acoustically similar, yet text-based metrics would treat them as entirely unrelated. Prior studies have shown that NLP metrics correlate poorly with human preference on audio based open-ended tasks such as Automated Audio Captioning (AAC) \cite{fense}.

\noindent \textbf{Audio captioning metrics.} Recognizing the inadequacy of NLP metrics for audio-based tasks, several evaluation methods have been proposed specifically for audio captioning. FENSE \cite{fense} measures semantic similarity using Sentence-BERT embeddings, while incorporating a fluency penalty to down-rank grammatically incorrect responses. SPICE+ \cite{gontier2023spice+} and ACES \cite{wijngaard2023aces} further refine evaluation by parsing captions into structured representations or sound descriptors before comparing embeddings. MACE \cite{dixit2024mace} combines two complementary CLAP embedding similarity scores: the score between audio and candidate captions, and the score between candidate and reference captions, capturing both acoustic grounding and semantic alignment. While these audio-based metrics have shown better alignment with human preferences for audio captioning, they are fundamentally question-agnostic and only compare the response to the reference. This makes them ill-suited for AQA, where contextual understanding of the question is essential.

\noindent \textbf{LLM-based metrics.} 
In the text domain, recent research has explored the use of large language models (LLMs) as evaluators for open-ended generation. Metrics like GEMBA \cite{kocmi2023large} (for machine translation) and G-Eval \cite{liu2023g} (for summarization) use LLMs to score model outputs directly, leveraging their ability to reason semantically. For question answering, prior work \cite{zheng2023judging, lin2023llm} show that instruction-tuned LLMs can reliably assess the correctness of answers, particularly when provided with in-context examples. In vision-language settings, LAVE \cite{manas2024improving} demonstrates that LLMs can effectively evaluate Visual Question Answering (VQA) by conditioning on the image, question, and answer. However, no prior work has extended this to Audio Question Answering. In this work, we first create a benchmark to evaluate how well the existing metrics align with human preferences and then propose an LLM-based metric for open-ended AQA evaluation.

\section{Human Annotation Details}
\label{sec:human_annotation}
\begin{figure*}
    \centering
    \includegraphics[width=1\linewidth]{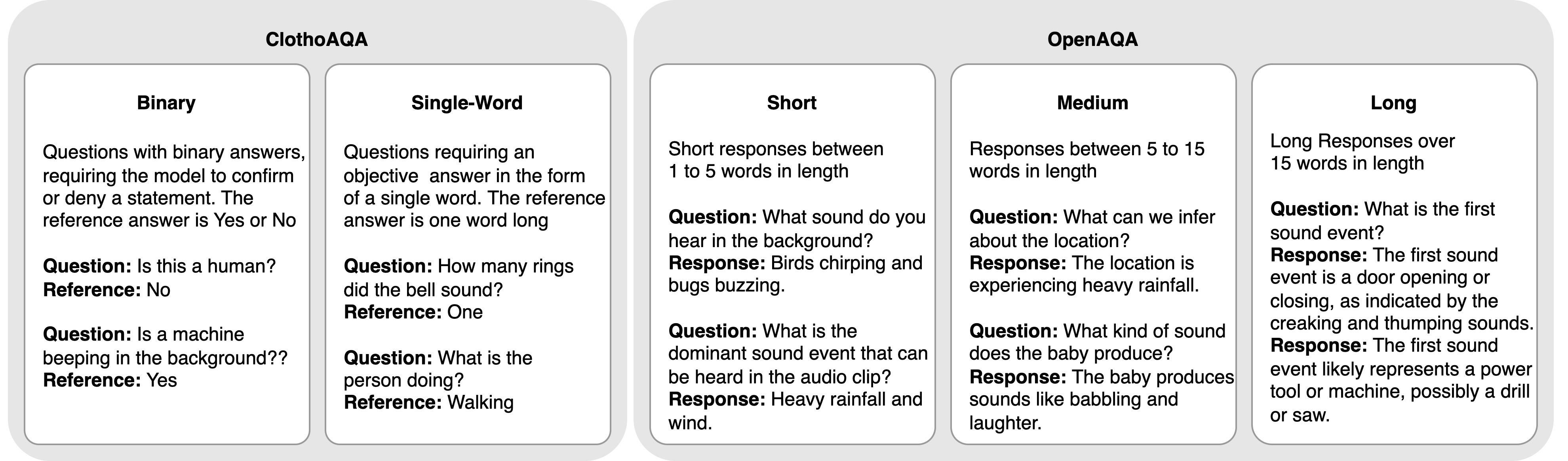}
    \vspace{-0.2 in}
    \caption{\textbf{Examples of AQEval question categories.} The figure showcases different types of questions and their reference answers from the AQEval dataset.}
    \vspace{-0.2 in}
    \label{fig:QTypes}
\end{figure*}

\subsection{Setup}
To collect human annotations, we utilized Amazon Mechanical Turk (MTurk) as the crowdsourcing platform. Each task was compensated with a reward of \$0.10. Workers were presented with audio clips and corresponding questions, as exemplified in Figure~\ref{fig:mturk_setup}. For each response, five independent annotations were collected to ensure reliability and reduce bias. Annotators were required to compare predicted responses against reference answers and classify them as either \textit{Correct} or \textit{Incorrect} based on the audio content.

\subsection{Annotator demographics}
To ensure high-quality annotations, we applied strict qualification requirements for MTurk workers:
\begin{itemize}[nosep]
    \item \textbf{HIT Approval Rate:} Workers were required to have a HIT approval rate of at least 98\%.
    \item \textbf{Location:} Workers were restricted to the United States.
    \item \textbf{Number of Approved HITs:} Workers needed to have completed more than 500 approved HITs.
\end{itemize}

\section{Qualitative examples}
\label{sec:qualitative}
Table~\ref{tab:qualitative_results} provides representative examples from the AQEval dataset, illustrating the question, reference answer, candidate answer, human score, and the scores from both a traditional metric (METEOR) and the proposed metric (AURA). These examples highlight the strengths of AURA in handling challenging cases by leveraging its ability to reason through an LLM and encode the underlying audio context effectively.

From the table, we observe that AURA consistently aligns with human judgments in cases where METEOR fails. For instance, in the first example, while METEOR assigns a low score due to lexical differences, AURA correctly identifies semantic equivalence between the candidate and reference answers. Similarly, in cases where candidate answers are more detailed or specific (e.g., ``aluminum foil'' instead of ``metal''), AURA captures this nuance and assigns a perfect score, matching human annotations. 

Conversely, METEOR often overestimates similarity in cases with shared lexical content but incorrect semantic meaning (e.g., ``dog barking'' instead of ``cat meowing''). Additionally, for ambiguous questions with multiple plausible answers (e.g., ``sheep'' vs. ``goat''), both AURA and human annotators assign partial credit, reflecting the inherent uncertainty in such tasks. These observations demonstrate AURA's robustness and its ability to outperform traditional metrics by reasoning at a deeper semantic level.

\begin{table*}[htp]
\centering
\begin{adjustbox}{width=\textwidth}
\begin{tabular}{>{\raggedright\arraybackslash}p{0.25\linewidth}>{\raggedright\arraybackslash}p{0.15\linewidth}>{\raggedright\arraybackslash}p{0.25\linewidth}>{\centering\arraybackslash}p{0.1\linewidth}>{\centering\arraybackslash}p{0.1\linewidth}>{\centering\arraybackslash}p{0.1\linewidth}>{\raggedright\arraybackslash}p{0.25\linewidth}}
\toprule

Question&  Reference Answer&  Candidate Answer &   Human&METEOR&  AURA & Observation\\ 
\midrule
Is there only one bird chirping?& no& No, multiple birds can be heard chirping in the background.&  1.000&0.104& 1.000& The candidate answer has the same meaning but has more details than the reference\\
\midrule
Can you identify the type of vehicle in the audio clip based on its acoustic feature?& It is a bus& The type of vehicle in the audio clip based on its acoustic feature is a bus.&  1.000&0.208& 1.000& The candidate answer has the same meaning but has more details than the reference\\
\midrule
What is the item that is rustling like paper?& metal& The item that is rustling like paper is likely aluminum foil.&  1.000&0.000& 1.000& The candidate answer is correct, only more specific\\
\midrule
What activity is taking place?& reading& Someone is flipping through pages of a book or newspaper.&  1.000&0.000& 1.000& The candidate answer is correct, only more specific\\
\midrule
What animal makes the sound?& sheep& The animal making sounds is a goat, as indicated by the bleating and baaing.&  0.500&0.000& 0.500& Many plausible answers for this question given the audio\\
\midrule
What type of ball is being struck?& tennis& a small, hard ball&  0.500&0.000& 0.500& Many plausible answers for this question given the audio\\
\midrule
What is the dominant sound event in the audio clip?& The dominant sound event is a cat meowing.& The dominant sound event is a dog barking&  0.000&0.502& 0.000& The common sequence of words leads to a high METEOR score for an incorrect answer\\

\bottomrule
\end{tabular}
\end{adjustbox}
\caption{Examples where AURA correctly evaluates responses that traditional metrics fail at.}
\label{tab:qualitative_results}
\end{table*}

\section{Prompts}
\label{sec:prompts}

The AURA metric relies on two structured prompts to guide the LLM's reasoning process for its scoring and audio entailment components. The first prompt, shown in Figure \ref{fig:hypothesis_prompt}, is designed for hypothesis generation. It uses few-shot in-context learning to instruct the LLM to synthesize the question and candidate response into a concise, declarative hypothesis. This reformulated statement is essential for the downstream audio entailment check, as it creates a factual claim that can be directly compared against the audio content. The second prompt, detailed in Figure \ref{fig:evaluation_prompt}, is used for answer evaluation. It leverages both few-shot examples (shown in Figure \ref{fig:evaluation_examples}) and a chain-of-thought approach, instructing the model to first generate a rationale explaining its reasoning before assigning a final correctness score. This design ensures that the LLM's evaluations are not only accurate but also transparent and grounded in explicit reasoning, which, as shown in our ablations, improves correlation with human judgment.

\begin{figure*}[htbp]
\centering
\begin{tcolorbox}[colback=gray!10, colframe=gray!50, boxrule=0.5pt, arc=3pt, left=6pt, right=6pt, top=6pt, bottom=6pt]
\footnotesize
\begin{verbatim}
Given the following question and response, generate a hypothesis that combines the information from 
both. The hypothesis should be a clear, standalone statement that can be evaluated against audio
content. Ensure the hypothesis captures all relevant details, especially when the response is 
complex or detailed.

Question: Is there a dog barking in the audio?
Response: Yes, there is a dog barking loudly in the background, and you can hear the dog's paws tapping
on the ground.
Hypothesis: A dog is barking loudly in the background, and its paws can be heard tapping on the ground.

Question: What kind of vehicle can you hear?
Response: I can hear a motorcycle engine revving, and there is a high-pitched whine indicating it's a
sport motorcycle.
Hypothesis: A sport motorcycle engine is revving, producing a high-pitched whine.

Question: Are people talking?
Response: No, there are no voices or speech in the audio.
Hypothesis: No people are talking in the audio

Now, for the following:
Question: {question}
Response: {response}
Hypothesis:

Generate a hypothesis that represents what should be true in the audio based on this question-response 
pair. Return only the hypothesis statement without any prefixes or explanations.
\end{verbatim}
\end{tcolorbox}
\caption{\textbf{Hypothesis generation prompt.} This prompt instructs the LLM to reformat a question-response pair into a single, declarative statement. The generated hypothesis is then used by the audio entailment module to verify if the claim is supported by the audio content evaluation.}
\label{fig:hypothesis_prompt}
\end{figure*}

\begin{figure*}[htbp]
\centering
\begin{tcolorbox}[colback=gray!10, colframe=gray!50, boxrule=0.5pt, arc=3pt, left=6pt, right=6pt, top=6pt, bottom=6pt]
\footnotesize
\begin{verbatim}
You are given a question, a reference answer written by experts, and a candidate answer. Please rate the
accuracy of the candidate answer for the question considering the reference answer.

Use a scale of 1-3, with 1 indicating an incorrect or irrelevant answer, 2 indicating an ambiguous or 
incomplete answer, and 3 indicating a correct answer.

Here are some examples:

Question: {example["question"]}
Reference answer: {example["reference_answer"]}
Candidate answer: {example["candidate_answer"]}
Output: The candidate answer is {example["reason"]} and therefore the score is {example["score"]}.

Now evaluate the following: Here is the question: {question}, the reference answer is: 
{reference_answer}, and the candidate answer is: {predicted_answer}. Give the rationale 
before rating. Format your response as a dictionary with a key "score", value either 
1, 2 or 3 and a key "reason" with a string value explaining your assessment.
\end{verbatim}
\end{tcolorbox}
\caption{\textbf{LLM Scoring prompt.} It instructs the LLM to provide a rationale before assigning a score of 1 (incorrect), 2 (ambiguous), or 3 (correct) to the candidate answer based on its alignment with the reference.}
\label{fig:evaluation_prompt}
\end{figure*}

\begin{figure*}[htbp]
\centering
\begin{tcolorbox}[colback=gray!10, colframe=gray!50, boxrule=0.5pt, arc=3pt, left=6pt, right=6pt, top=6pt, bottom=6pt,]
\footnotesize
\begin{verbatim}
{ # correct 1
    "question": "Is the source of the sound indoors?",
    "reference_answer": "yes",
    "candidate_answer": "Yes, the audio suggests that the source is inside a room or building, as 
    indicated by the consistent bouncing sounds and lack of external noise like wind.",
    "score": 3,
    "reason": "The candidate answer is correct because it is equivalent to the reference answer"
},
{ # correct 2
    "question": "What can you infer about the weather conditions when this audio was recorded?",
    "reference_answer": "There was wind blowing during the recording.",
    "candidate_answer": "The presence of wind noise suggests that it might be a breezy or windy day, 
    possibly in an outdoor setting near water.",
    "score": 3,
    "reason": "The candidate answer is correct because it is equivalent to the reference answer"
},
{ # correct 3
    "question": "Is there a dog barking in the audio?",
    "reference_answer": "yes",
    "candidate_answer": "there is a dog barking",
    "score": 3,
    "reason": "The candidate answer is correct because it is equivalent to the reference answer"
},
{ # ambiguous 1
    "question": "Does the motor stop?",
    "reference_answer": "No.",
    "candidate_answer": "The audio does not provide information about whether the motor stops or 
    continues running",
    "score": 2,
    "reason": "The candidate answer is ambiguous"
},
{ # ambiguous 2
    "question": "is the sound scary?",
    "reference_answer": "no",
    "candidate_answer": "The continuous roar of a running engine and the thunderstorm might be 
    perceived as intense or dramatic, but it depends on individual perceptions and experiences.",
    "score": 2,
    "reason": "The candidate answer is ambiguous"
},
{ # ambiguous 3
    "question": "What acoustic features in the audio clip indicate that a machine is running?",
    "reference_answer": "The loud sound in the audio clip indicates that a machine is running.",
    "candidate_answer": "The continuous humming sound indicates the operation of a machine, likely an 
    engine or motor.",
    "score": 2,
    "reason": "The candidate answer is ambiguous because while it is one possible answer, it is 
    incomplete"
},
{ # incorrect 1
    "question": "What kind of sound do you hear in the background?",
    "reference_answer": "Bird chirping",
    "candidate_answer": "The background sound is a vehicle passing by, which suggests that the scene 
    takes place near a road or path.",
    "score": 1,
    "reason": "The candidate answer is incorrect because it contradicts the reference answer"
},
{ # incorrect 2
    "question": "Are people talking?",
    "reference_answer": "no",
    "candidate_answer": "Yes, there is a faint sound of human speech in the audio.",
    "score": 1,
    "reason": "The candidate answer is incorrect because it contradicts the reference answer"
},
{ # incorrect 3
    "question": "Is someone coughing?",
    "reference_answer": "no",
    "candidate_answer": "Yes, a person is coughing in the audio.",
    "score": 1,
    "reason": "The candidate answer is incorrect because it contradicts the reference answer"
}
\end{verbatim}
\end{tcolorbox}
\caption{\textbf{Few-shot examples.} Demonstrations used within the evaluation prompt (Figure \ref{fig:evaluation_prompt})}
\label{fig:evaluation_examples}
\end{figure*}

\end{document}